# Electron Track Reconstruction and Improved Modulation for Photoelectric X-ray Polarimetry

Tenglin Li [a,b], Ming Zeng [a,b] *, Hua Feng [a,b,c], Jirong Cang [a,b], Hong Li [a,b,c], Heng Zhang [a,b,c], Zhi Zeng [a,b], Jianping Cheng [a,b], Hao Ma [a,b], Yinong Liu [a,b]

[a] *Department of Engineering Physics, Tsinghua University, Beijing 100084, China*

[b] *Key Laboratory of Particle & Radiation Imaging (Tsinghua University), Ministry of Education, China*

[c] *Center for Astrophysics, Tsinghua University, Beijing 100084, China*

**Abstract**

The key to photoelectric X-ray polarimetry is the determination of the emission direction of photoelectrons. Because of the low mass of an electron, the ionisation trajectory is not straight and the useful information needed for polarimetry is stored mostly in the initial part of the track where less energy is deposited. We present a new algorithm, based on the shortest path problem in graph theory, to reconstruct the 2D electron track from the measured image that is blurred due to transversal diffusion along drift and multiplication in the gas chamber. Compared with previous methods based on moment analysis, this algorithm allows us to identify the photoelectric interaction point more accurately and precisely for complicated tracks resulting from high energy photons or low pressure chambers. This leads to a better position resolution and a higher degree of modulation toward high energy X-rays. The new algorithm is justified using simulations and measurements with the gas pixel detector (GPD), and it should also work for other polarimetric techniques such as a time projection chamber (TPC). As the improvement is restricted in the high energy band, this new algorithm shows limited improvement for the sensitivity of GPD polarimeters, but it may have a larger potential for low-pressure TPC polarimeters.

# 1 Introduction

Sensitive X-ray polarimetry in astrophysics based on the photoelectric effect has become possible in recent years with the development of high-resolution micro-pattern gas detectors [1, 2]. The azimuthal distribution of the photoelectron direction is dependent on the polarisation of the X-ray. The objective of these gas polarimeters is to measure the 2D photoelectron directions on the plane perpendicular to the incident X-rays. The sensitivity of the polarimeter depends on how accurately and precisely the direction can be measured. Thus, the algorithm to reconstruct the photoelectron emission direction from the measured track image is essential.

* Corresponding author, *E-mail*: zengming@tsinghua.edu.cn

Previously, the reconstruction for the electron emission direction was achieved based on the moment analysis of the track image. The principles of the algorithm were introduced in Ref. [3] and the reconstructions of the interaction point and emission direction were later improved [4]. Here we summarise the basic steps of the moment analysis based algorithm, and more details can be found in Section 2.2.1 in Ref. [4]. The second moments of the entire image (charge deposition) in all directions with respect to the barycentre are first calculated. The elongation of the image can be found along the direction $\Phi$ where the second moment is maximised. The third moment of the image along the direction $\Phi$ is then calculated to determine which part of the image along the elongation axis contains more charges. The end with less charge is regarded as the initial part of the track. Pixels around the photoelectric interaction point are selected from those between a smaller and a larger radius (normalised to the maximum second moment) from the barycentre along the direction with less charge, and the interaction point is estimated as their centre of mass. Once the interaction point is found, the emission angle can be derived as the direction of the new maximum second moment of a distance-weighted charge map around it.

The most important and challenging part of the direction reconstruction is to determine how to locate the interaction point accurately and precisely. However, the algorithm discussed above may fail for complicated tracks, as it simplifies the track image into elliptical distributions of charges. For example, if the photoelectron moves back and forth and does not stop at the very end, the algorithm may misidentify the location of the interaction point. For a gas pixel detector (GPD) filled with 0.8 atm dimethyl ether (DME) [1, 5], the current algorithm produces reasonable modulations at energies below ~7 keV, but it is not optimal at energies above due to complications of the electron tracks. The situation may be even worse for the time projection polarimeters, as some of them are filled with low-pressure (0.25 atm) gas [6].

In this paper, we propose a new algorithm to locate the interaction point via a reconstruction of the full electron track. The algorithm is described in Section 2. The results tested with the GPD polarimeter using simulated and measured data are discussed in Section 3. The conclusion is summarised in Section 4.

## 2 The track reconstruction algorithm

In this paper, the algorithm is demonstrated and tested with simulated and measured data using the GPD polarimeter, for which the detector structure and configuration can be found in Ref. [5]. The sensitive gas volume is 1 cm thick (the drift distance), which stands above the GEM foil and is filled with pure dimethyl ether (DME) at 0.8 atm. The image has hexagonal pixels with a pitch of 50 μm and the full width at half maximum (FWHM) for the total transversal diffusion is on the order of 150 μm.

### 2.1 Track and interaction point reconstruction

We used a simulated event generated by a 15 keV X-ray photon to illustrate the track reconstruction algorithm. The photon deposits all its energy by photoelectric effect, and the photoelectron leaves behind a track of charge via ionisation (see Fig. 1a for a 2D projection of the track on the readout plane, simulated using GEANT4 [7]). Secondary electrons created by the photoelectron will drift along the electric field toward the readout plane and be multiplied when they go through the gas electron multiplier (GEM). This process will introduce transversal diffusion and dilute the charge image by a factor of $\sigma_{drift}$ and $\sigma_{multi}$, respectively, above the GEM and below the GEM. The value of $\sigma_{drift}$ is distance dependent ($\sigma_{drift} = 70\sqrt{d}$ μm, where $d$ (cm) is the drift distance) and $\sigma_{multi}$ is fixed at 35 μm; the standard deviations of the two-dimensional Gaussian transversal diffusion function can be described as $\sigma_x = \sigma_y = \sqrt{\sigma_{drift}^2 + \sigma_{multi}^2}$. The electrons are then collected on the hexagonal pixels of the readout chip when they approach the detector plane, on top of the readout noise (Fig. 1b). Our simulation package can produce images and statistical results that are highly consistent with the real measurements [5], justifying the use of the simulations for testing the new algorithm in this paper.

Because of the presence of a noise threshold, the measured charge distribution (image) may be broken into several disconnected segments (or clusters of points). This is particularly true at higher energies where the specific linear ionization is minimal before the Bragg peak where the kinetic energy is lower. The example illustrated in Fig. 1c consists of three individual clusters, denoted as A, B, and C. The *clustering algorithm* is used to identify individual clusters and the neighbouring ones are connected with each other via their nearest pixels if they are closer than a certain threshold (0.25 mm in this case). Clusters of less than 10 pixels are discarded as they may be caused by noise or other effects such as absorption of re-emitted characteristic X-rays in the gas. The 2D image can then be used for track reconstruction, following the steps described below.

1. **Connecting points** (Fig. 1c). In graph theory, the *points* are mathematical abstractions corresponding to interconnected objects and the *lines* describe the connectivity between the points. In this paper, points are defined as the pixelated energy depositions and two points are connected with a line if they are next to each other (i.e., the space between two points equals 50 μm).

2. **Calculating the primary path** (Fig. 1d). According to the shortest path problems in graph theory [8], we can calculate the *shortest path* given by any two points in the graph. The length of a path in this paper is defined as the actual spatial length of it. Amongst all the shortest paths between every two points, we select the longest one, defined as the *primary path*, which is a rough estimate of the track (see Fig. 1d). If the longest path is not unique, the one with the minimum deflection is selected.

3. **Spatial filtering and track reconstruction** (Fig. 1e). The primary path found above often traces the edge of the track pixels. To make it trace the centroid of the track, we apply a spatial energy filter to smooth the polyline: every point in the primary path is replaced by the charge barycentre within a certain radius around it. Thus, a new track (called a *reconstructed path*, see Fig. 1e) is obtained, which is smoother and closer to the centre of the ionisation track than the primary path. The filter radius (0.22 mm in our case) is related to the size of the total transversal diffusion, to match the actual width of the track.

4. **Interaction point reconstruction (Fig. 1e).** By calculating the sum of the charge

deposition around each endpoint of the reconstructed path, the one with less charge is determined as the photoelectric interaction point, and the other one is the Bragg peak.

With these steps, the reconstructed path and interaction point are obtained. A comparison of the reconstructed path and the original ionisation track is shown in Fig. 1f. It is obvious that the new algorithm can provide a reasonable reconstruction for the track and an accurate and precise estimate of the interaction point.

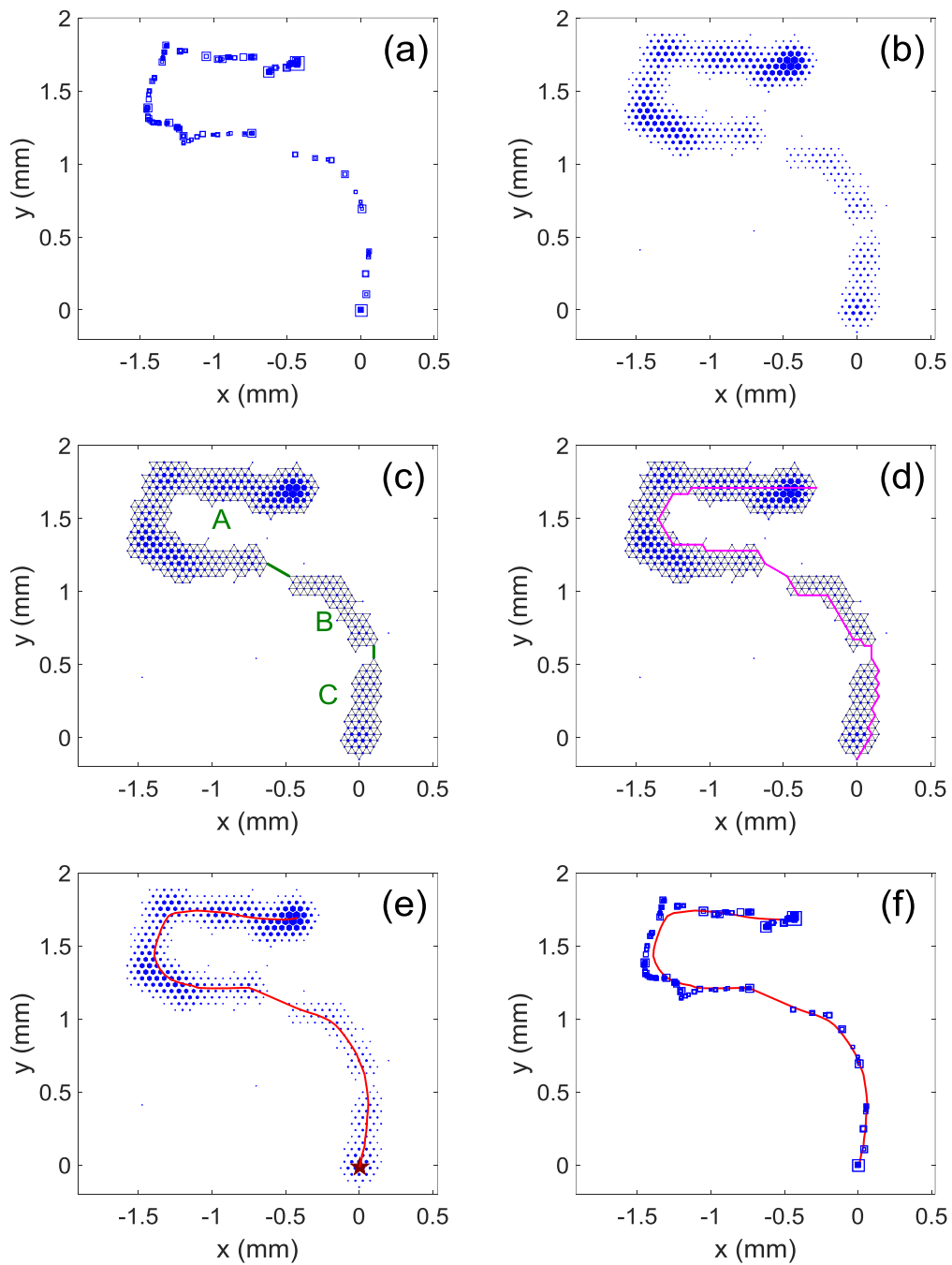

Fig. 1. Illustration of the track reconstruction algorithm. **(a)** Simulated charge distribution produced by a 15 keV X-ray projected onto the plane perpendicular to the X-ray direction. The marker size is proportional to the energy deposition (similarly hereafter). **(b)** Image obtained by the readout ASIC of the GPD detector. **(c)** Neighbouring clusters and neighbouring points are connected with thick and thin lines, respectively. **(d)** For all shortest paths between every two points in the graph, the longest one is defined as the primary path (thick line). **(e)** The reconstructed path (thick line) is derived from the primary path after spatial energy filtering. The end with less charge depositions is determined as the interaction point (star). **(f)** Comparison between the reconstructed path and the initial charge distribution.

## 2.2 Emission angle reconstruction

Once the interaction point is estimated, the 2D emission direction of the photoelectron is calculated as the direction of the maximum second moment of a distance-weighted map around the interaction point. Adopted from Ref. [4], the weight $W(d_{ip})$ is defined as:

$$W(d_{ip}) = \exp(-d_{ip}/w),$$

where $w$ is a constant derived from Monte Carlo simulations (0.05 mm in this case) and $d_{ip}$ is the distance from each pixel to the estimated interaction point. The emission angle is then computed as the direction of the maximum second moment of the weighted map.

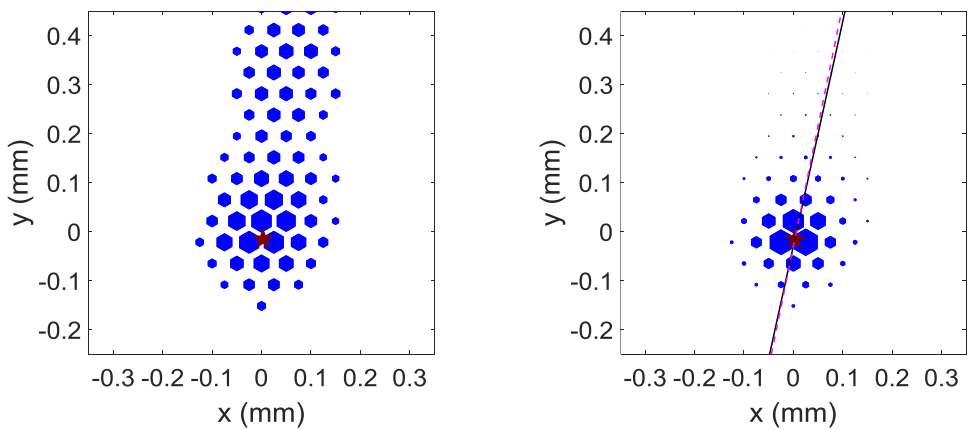

Fig. 2. The detected map (left) and the weighted map (right) around the reconstructed interaction point (star). The reconstructed emission direction (dashed) approximates the true emission direction (solid).

## 3  Results and Discussion

We simulated GPD detected events with polarised and unpolarised X-rays using the GEANT4 package. The effects of diffusion during transportation and multiplication for secondary electrons in the gas chamber were implemented based on simulations with the GARFIELD package[9]. It is two-fold that the simulation is to some extent a good representation of the real measurement: the simulated and measured images are not distinguishable even with trained eyes; the simulated and measured modulations are well consistent with each other within errors[5]. For polarised X-rays, $10^5$ events are simulated at each energy. For unpolarised X-rays, $10^6$ events are generated to see the possible systematics of small amplitude.

For each data set, both the old algorithm based on the moment analysis and the new algorithm based on track reconstruction are applied and compared. The difference between the old and the new algorithm is how to locate the interaction point, and after that the two algorithms become identical in calculating the emission direction from a weighted charge map.

All the events are divided into two subsets according to the eccentricity, defined as the maximum to minimum ratio of the second moment. For events with an eccentricity larger than 2, both the old and new algorithms are adopted for interaction point reconstruction. For those having an eccentricity smaller than 2, the emission angles are simply calculated as the direction of the maximum second moment. The fraction of low eccentric (less than 2) events ranges from ~26% at 5 keV, ~14% at 6 keV, to only ~3% at 15 keV.

### 3.1 Deviation of the reconstructed interaction point

We analysed the simulated track images, compared the interaction points computed with the two algorithms and also compared with interaction points known a priori since it was imposed in the simulation. Fig. 3 shows the 2D distribution of the reconstructed point around the true point, and Fig. 4 shows the distribution of the distance between them. Compared with the moment analysis based algorithm, the misidentification rate for the new algorithm is lower and the reconstruction is concentrated around the true interaction point. This is particularly remarkable for high-energy or complicated tracks. We visually checked the instances of misidentification for both algorithms and found that for the new algorithm almost all the misidentifications happened in cases where human eyes could not identify the interaction point, e.g., for looped tracks. On the other hand, the moment analysis based algorithm often found an incorrect interaction point as per visual inspection (see Fig. 5).

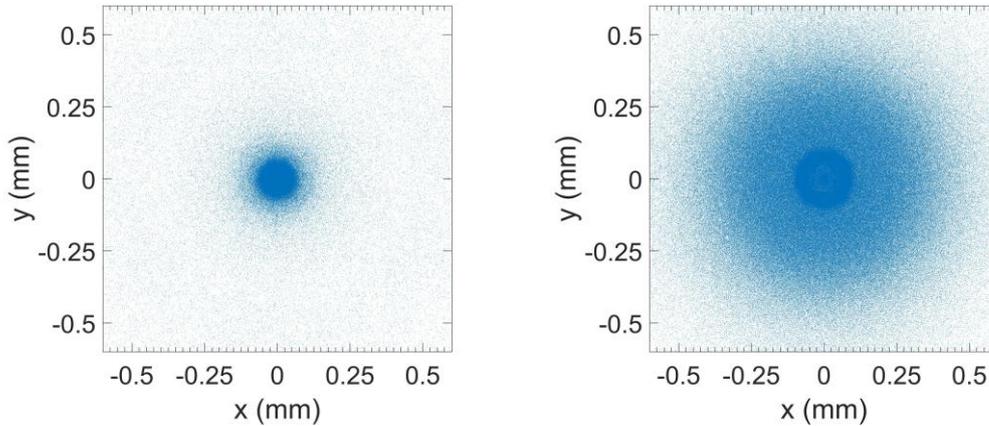

Fig. 3. Distribution of the reconstructed interaction point with respect to the true interaction point (moved to the origin) for the track reconstruction based algorithm (left) and the moment analysis based algorithm (right) for X-rays at 15 keV.

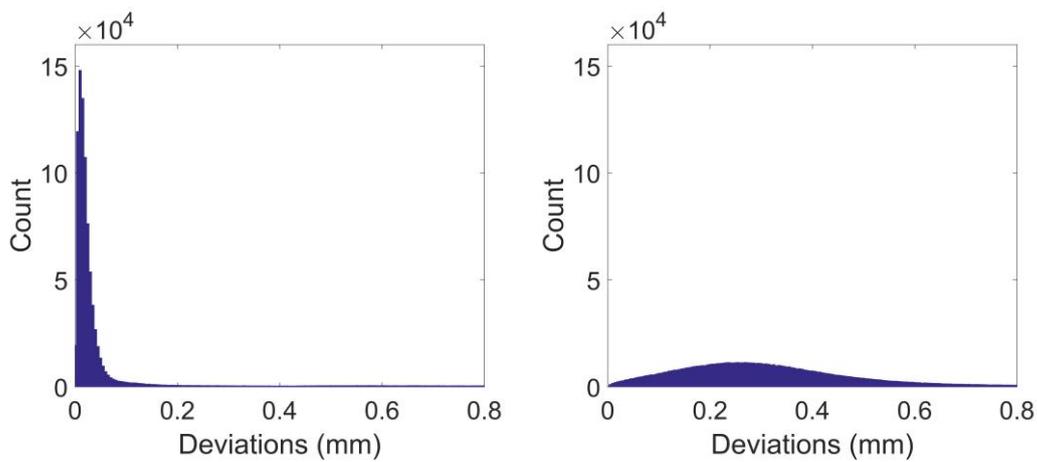

Fig. 4. Distribution of the distance between reconstructed and true interaction points for the track reconstruction based algorithm (left) and the moment analysis based algorithm (right).

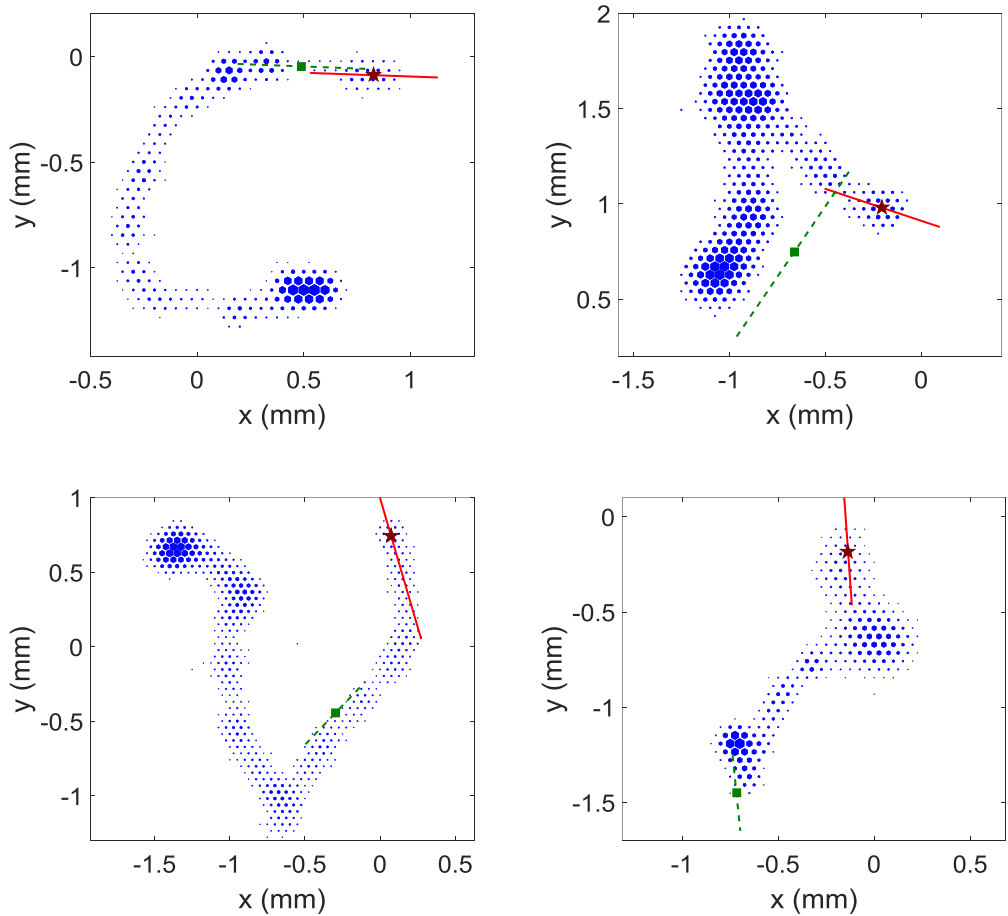

Fig. 5. Typical examples where the old algorithm fails but the new algorithm can successfully find the true interaction point. The star and solid line indicate the reconstructed interaction point and the emission direction, respectively, found by the new algorithm, while the square and dashed line indicate those for the old, moment analysis based algorithm.

In addition to lowering of the degree of modulation, misidentification of the interaction point will also dilute the imaging resolution. The contribution to the half power diameter (HPD) of the point spread function by this factor is computed and shown in Fig. 6 at different energies. The position resolution based upon the reconstructed interaction point derived from the old algorithm is as accurate as from the new one at energies below 6 keV (we note that the computed position resolution at 4.5 keV is consistent with that reported in Ref. [10]).

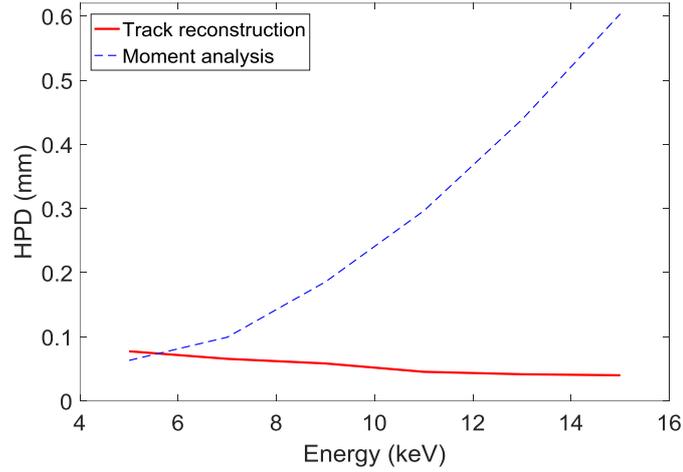

Fig. 6. Position resolution of the GPD in terms of HPD derived from the old and new algorithm, respectively, as a function of photon energy.

### 3.2 Modulation factors

The modulation curves in response to fully polarised X-rays constructed from emission angles based on the new algorithm at different energies, from 5 to 15 keV, are shown in Fig. 7. They are highly consistent with sinusoidal curves. The modulation factors versus energy are shown in Fig. 8 and listed in Table 1. The new algorithm only manifests advantages over the old, moment analysis based algorithm at energies of 7 keV and above. For comparison, the modulation factors calculated with known interaction points from the simulation data are shown in Fig. 8.

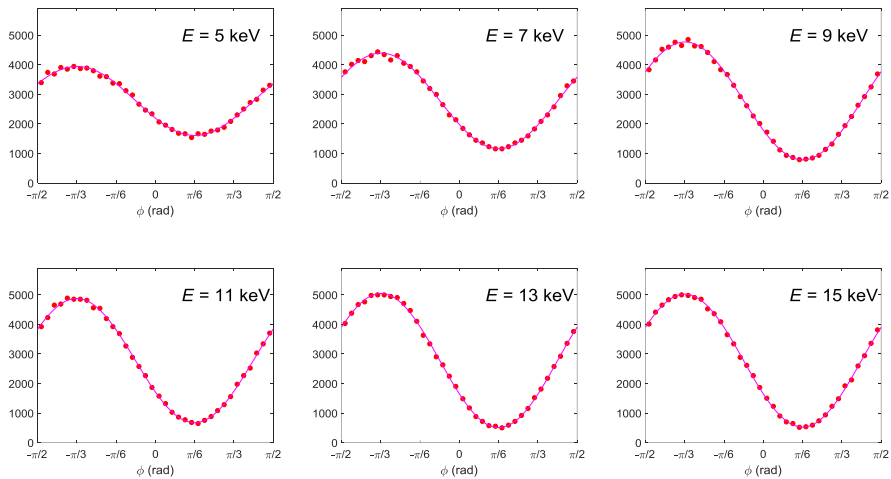

Fig. 7. Modulation curve obtained with the new algorithm at different energies from the simulated data. The error bars are smaller than the point.

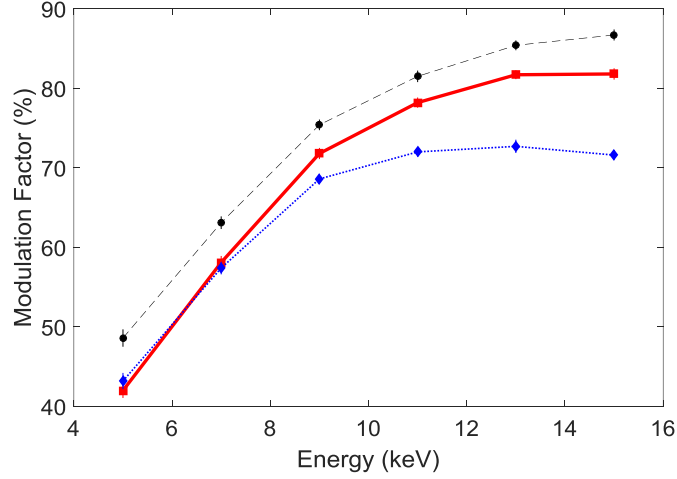

Fig. 8. Modulation factors versus energy derived from the moment analysis based algorithm (dotted line), track reconstruction based algorithm (solid line), and with known interaction points (dashed line).

Table 1. The modulation factors obtained by both algorithms

| Energy ($E$) | 5 keV | 7 keV | 9 keV | 11 keV | 13 keV | 15 keV |
| --- | --- | --- | --- | --- | --- | --- |
| $\mu$ from track reconstruction (%) | 42.0 ± 0.9 | 58.1 ± 0.8 | 71.8 ± 0.7 | 78.2 ± 0.6 | 81.7 ± 0.6 | 81.8 ± 0.7 |
| $\mu$ from moment analysis (%) | 43.2 ± 1.0 | 57.4 ± 0.8 | 68.6 ± 0.6 | 72 ± 0.6 | 72.7 ± 0.8 | 71.6 ± 0.6 |

## 3.3 Modulation factors with experimental data

We also tested the new algorithm with the real measurements, as shown in Fig. 9. The data were adopted from Ref. [5], for fully polarised X-rays at 5.33, 6.09, and 7.49 keV, respectively. To have a direct comparison with the results quoted in Ref. [5], which discarded 25% of the events with low eccentricity, we adopted the same cut for both algorithms. The results from the measurements are consistent with those from the simulation within errors, as expected. As the measurements are not available at energies above 8 keV, we cannot directly test the validity of the algorithm further for complicated tracks. The good agreement between the simulation and measurement at low energies justifies the use of the simulation at high energies.

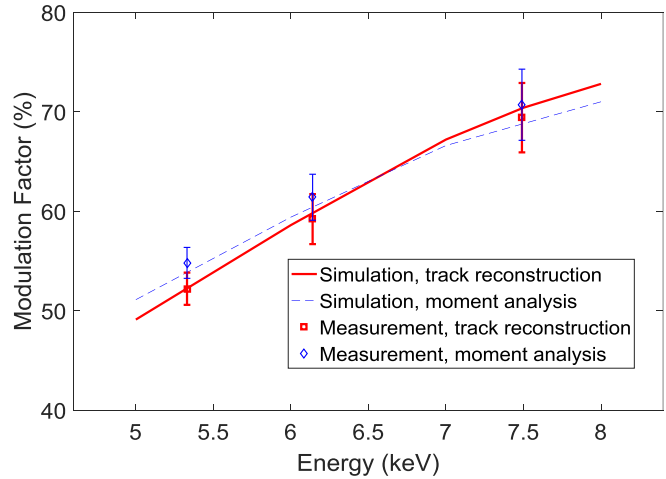

Fig. 9. Comparisons of the modulation factors obtained from both measurements and simulations for both algorithms.

### 3.4 Test with unpolarised data

To test if the new algorithm would produce spurious modulations, we applied it to simulated unpolarised data at 6 and 15 keV. A million events were generated at each energy, corresponding to a minimum detectable modulation of 0.43% at 99% confidence level (a chance of 1% to get a measurement higher than 0.43% given a zero polarised source). Both tests resulted in a null detection within errors, with a modulation of 0.35% ± 0.31% at 6 keV and 0.15% ± 0.26% at 15 keV. We noted that the true distribution of the emission angle for this simulated data set had a modulation of 0.11% ± 0.27% at 6 keV and 0.09% ± 0.30% at 15 keV. Thus, the new algorithm does not introduce detectable systematics.

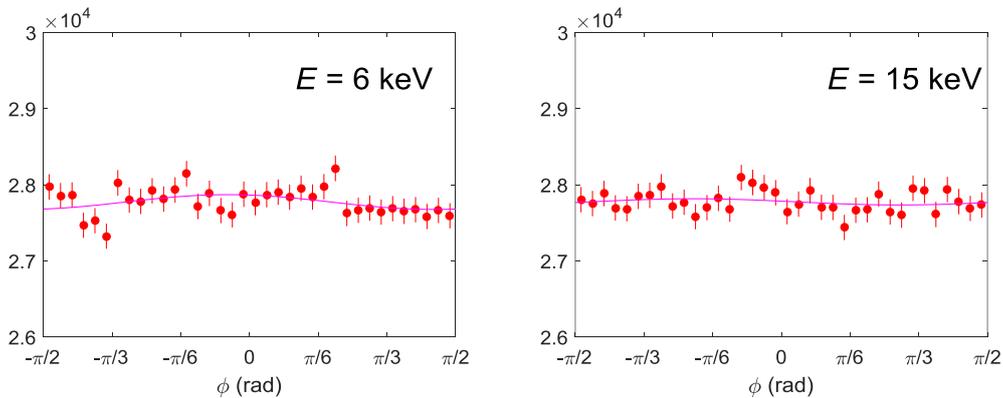

Fig. 10. Modulation curves and best-fit $\cos^2$ functions for unpolarised X-rays at 6 keV and 15 keV, respectively. The residual modulation was found to be 0.35% ± 0.31% at 6 keV and 0.15% ± 0.26% at 15 keV.

3.5 **Computational cost**

We compared the computational cost of the new and old algorithms by processing a million simulated events at 6 keV through the full chain. On the same computer (Intel Core i7-3770, 3.40 GHz), the moment analysis based algorithm takes about 36 minutes, while the track reconstruction based algorithm (a MATLAB script without optimization) takes about 180 minutes. The increase in the computational complexity of the new algorithm is still acceptable for practical use.

# 4  Conclusion and perspective

In this paper, we propose a new algorithm that allows us to reconstruct the 2D photoelectron track in the gas chamber produced by the absorption of an X-ray photon. Compared with previous algorithms that are based on the moment analysis, this algorithm is more effective and robust for complicated tracks resulting from the high-energy photons or low-pressure chambers. The algorithm can help locate the photoelectric interaction point more accurately (less misidentification) and precisely (smaller scatter) than previous algorithms, resulting in a better position resolution and a higher modulation factor. Importantly, the new algorithm does not introduce detectable residual modulation or systematics, which, for bright sources, determines the limiting sensitivity of the detector.

However, the improvement of the sensitivity for the GPD by this new algorithm is limited. The sensitivity of a polarimeter can be described as the minimum detectable polarization (MDP), which is inversely proportional to the modulation factor. Due to the fast (power-law) decline of the source flux and effective area with increasing energy, one needs a large exposure time ($10^5$ - $10^6$ seconds) to collect a great number of photons ($10^5$ - $10^6$) needed for sensitive polarimetry according to the current design of IXPE, XIPE, or eXTP at energies of 8 keV and above, even for the brightest X-ray sources on the sky (e.g., the Crab nebula). The effective modulation factor weighted by the measured energy spectrum at energies above 8 keV using the new algorithm is only 5% better than using the old one. In other words, for those future X-ray polarimetric telescopes with the GPD as the focal plane detector, the new algorithm can offer an improvement only for very few sources at a limited level.

On the other hand, the new algorithm may be more useful for low-pressure TPC type polarimeters, in which, compared with GPD polarimeters, the electron track is longer and more complicated at the same energy and the detection efficiency is higher at high energies.

For the emission angle estimate, we still adopted the previous method, which calculates the maximum second moment of a weighted map around the interaction point. In principle, the emission angle can be obtained directly from the reconstructed path. We have tried various methods but all resulted in spurious modulations (deviated from a sinusoidal curve). This should be further investigated in the future.

## Acknowledgements

We thank the anonymous referees for useful comments that have improved the paper. This work was supported by the National Natural Science Foundation of China (Grant Nos. 11175099, 11305093 and 11633003), the Tsinghua University Initiative Scientific Research Program (Grant Nos. 2011Z07131, 2014Z21016 and 20131089244), and the National Program on Key Research and Development Project (Grant No. 2016YFA040080X).